\documentclass{article}
\usepackage[a4paper, margin=2.5cm]{geometry}
\usepackage[utf8]{inputenc}
\usepackage[T1]{fontenc}
\usepackage{lmodern}
\usepackage{amsmath}
\usepackage{amsthm}
\usepackage{amsfonts}
\usepackage{amssymb}
\usepackage{mathtools}
\usepackage{paralist}
\usepackage[shortlabels]{enumitem}
\usepackage{setspace}
\usepackage{booktabs}
\usepackage{caption}
\usepackage{longtable}
\usepackage{bm}
\usepackage{todonotes}
\usepackage{subfig}
\usepackage{url}
\usepackage{textcomp}
\usepackage{hyperref}
\usepackage{wrapfig}

\usepackage{bbold}
\usepackage[normalem]{ulem}

\newcommand{\note}[1]{} 
\newcommand{\source}[1]{}
\newlength{\mywidth}
\title{Has EU accession boosted patents performance in the EU-13? – A critical evaluation using causal impact analysis with Bayesian structural time-series models} 
\author{Agnieszka Kleszcz\footnote{E-mail:akleszcz@agh.edu.pl, Jan Kochanowski University of Kielce, Kielce, Poland} \footnotemark[2], 
Krzysztof Rusek\footnote{E-mail:krusek@agh.edu.pl, AGH University of Science and Technology, Krakow, Poland}
}
\date{}

\begin{document}
\footnotetext[1]{This work has been submitted to the Central European Journal of Economic Modelling and Econometrics and is under review process}
\maketitle
\begin{abstract}
Nowadays innovation is one of the main determinants of economic development. 
Patents are a key measure of innovation output, as patent indicators reflect the inventive performance of countries, technologies and firms.
This paper provides new insights on the causal effects of the enlargement of the European Union (EU) by investigating the patents performance within the new EU member states (EU-13). %
The empirical results based on data collected from the OECD database from 1985 – 2017 and causal impact using a Bayesian structural time-series model (proposed by Google) point towards a conclusion that joining the EU has had a significant impact on patents performance in Romania, Estonia, Poland, Czech Republic, Croatia and Lithuania, although in the latter two countries the impact was negative. 
For the rest of the EU-13 countries there is no significant effect on patent performance. Whether the EU accession effect is significant or not, the EU-13 are far behind the EU-15 (countries which entered the EU before 2004) in terms of patent performance. The majority of patents (98.66\%) are assigned to the EU-15, with just 1.34\% of assignees belonging to the EU-13.

\end{abstract}

\textbf{Keywords}: causal analysis, European Union, patents, innovations, Bayesian structural time-series models

\textbf{JEL Classification}: O340, C11, C32, O52 


\section{Introduction}
According to the European Patent Office~\cite{EPO} a patent is \emph{``legal title that gives inventors the right, for a limited period (usually 20 years), to prevent others from making, using or selling their invention without their permission in the countries for which the patent has been granted''}. 

In an increasingly knowledge‐driven economy, society invariably needs creative or inventive ideas or concepts to improve existing features, add useful new features to products or develop new products. 
European patent application statistics from 2019 (\cite{EPO2}) show the top patent technology fields were digital communication, medical technology and computer technology- all the highly influential and fast-growing areas of technology. 

In this rapidly changing environment patents are a long-term investment.
The patent applicant can commercialize the invention at any point during the time of patent protection, either through developing products or services incorporating the patented technology or by licensing it to others. 
This means patents play an increasingly important role in innovation and economic performance. 
There is evidence that patenting in general contributes positively to financial performance.
The number of patents is quantitative output of technologically successful R\&D activities and can be seen as one of the key-components of a country’s innovation \cite{maresch2016patents}.

In the literature (e.g. \cite{Lee2020}), innovation is presented as the most important factor in achieving economic and employment growth and one of the main determinants of economic development in modern societies. 
Investing in research is considered essential for achieving smart, sustainable and inclusive growth and jobs in Europe. Investment protects R\&D activities by approval of patents rights and this can make the patent a risk reducing factor for an investor. Given the strong connection between innovations and patents, we may expect that less innovative countries will be lagging behind the highly innovative ones in terms of total patent number.

Despite the efforts of European Union (EU) policy makers, disparities exist between the EU-15 (countries which entered the EU before 2004) and EU-13 (countries that joined the EU in and after 2004) in various ways as highlighted by many authors e.g. \cite{Filippetti2013, Makkonen2016, Pazour2020}. 
Among other things, the EU is also polarized in terms of innovations.
According to ~\cite{Hollanders2019} data from the European Innovation Scoreboard suggest that the EU-13 lags behind the EU-15. 
As showed by \cite{Kleszcz2020}, the headquarters of the EU’s top 1,000 R\&D investors from the ICT sector are located in 16 countries, with an overwhelming concentration in EU-15 countries.
Furthermore,  organisations based in the EU-13 have benefited less from their participation in the European Framework Programmes (FPs) than organisations from the EU-15 \cite{Fresco2015}.
Most of the time, the EU-13 can be found at the lower end of participation rankings concerning Horizon 2020 \cite{Pazour2018}. 
The majority of EU nations with smaller research budgets are former communist countries in central and eastern Europe, which together with Cyprus and Malta joined the EU after 2004 as emphasized by \cite{Abbott2019}.

In recent years, there have been numerous academic and policy debates on the delivery mechanisms of EU funds in member states. Accessing large amounts of funding has often been seen as one of the main benefits of EU accession. It was hoped that European funds like Structural Funds, direct payments in agriculture, rural development and others would contribute towards reducing disparities between regions. Economists and regional geographers have been very interested in studying the effects of European transfers on national and regional/local economies e.g.~\cite{Neculai2021}. 
In various scientific literature and reports analysis can be found assessing the impact of EU membership on different aspects like political changes, economic reforms, political values or foreign policies e.g.  \cite{van2021convergence,Nitoiu2020,baaseconomic,Felbermay}.\
Nevertheless, there are different views on the EU and its influence on member countries. Many of the Central and Eastern European countries surveyed positive views of the political union, however on the other hand there are plenty of Eurosceptics within the EU, as the UK Brexit referendum demonstrated. 
Hence it is crucial to reliably and quantitatively assess how membership impacts member states in many different aspects. Considering all aforementioned aspects of disparities between EU-13 and EU-15 leads to a natural question of: Has EU-13 accession boosted the innovation of these countries? Since the question is about the cause and an effect, a formal causal analysis is required to answer it. However, causal impact analysis of EU enlargement on the EU-13 countries’ innovations could be identified in very few scientific publications.
\cite{gyen_2018} investigates the relationship between innovation and EU membership using panel data on the firm level, while
\cite{Makkonen2016} studied scientific collaboration between ‘old’ and ‘new’ member states. Results from the latter paper are a direct motivation for this research as they show that the collaboration between the new and old member states has been affected by EU enlargement.

International scientific co-publication is just one of the components of innovation. The number of Patent Cooperation Treaty (PCT) patents applications per billion GDP is another  component of Summary Innovation Index ~\cite{Hollanders2019}.
In this paper, we used number of patents as one key measure of innovation output.
Despite the role of patents in innovation, no studies so far have attempted to examine the dynamics of patents performance in the EU-13 over time, particularly including the effects of EU-13 accession on patents performance. This study has tried to bridge that gap and analyse dynamical causal impact of EU accession on patents performance. 

Causal impact of EU accession on patents performance was assessed by utilizing Bayesian structural time-series models. The method generalises the widely used difference-in-differences approach to the time-series setting, by explicitly modelling the counterfactual of a time series observed both before and after the intervention. Hence, the research problem that guides the analysis is the assessment of the question: Has EU membership significantly increased the number of patents of the EU-13 countries.

Furthermore, the paper explores the total number of patents to highlight disparities between EU-13 and EU-15 countries.
The following research problems are addressed in the paper:
\begin{itemize}
    \item To quantify the effects of accession of EU-13, in particular find causalities if the EU-13 accession to EU has increased dynamics of patents performance in the EU-13.
    \item To analyse disparities between EU-13 and EU-15.
\end{itemize}

The main contribution of this paper is assessment of patents performance for EU-13 countries after accession to the EU using a new causal impact approach with Bayesian inference.
The method generalises the widely used difference-in-differences approach commonly used in related work to the time-series setting by explicitly modelling the counterfactual of a time series observed both before and after the intervention. 

The structure of this paper is as follows. In Section \ref{sec:Related.work} we provide a the review of the relevant literature on the patent analysis and different causal relationships in the innovation field. In Section \ref{sec:materialsmethods} we describe dataset with methodology. In Section \ref{sec:results} we report and discuss our empirical results. Section \ref{sec:conclusion}  summarizes and concludes with policy recommendations.

\section{Related work}
\label{sec:Related.work}

Intellectual property law governs technological innovation. In the literature, the field of innovations has been presented by \cite{Lee2020,Szopik-Depczynska2018} as the most important factor to achieve economic and employment growth. Patents are a key measure of innovation output, as patent indicators reflect the inventive performance of countries, technologies and firms. They are also used to track the level of diffusion of knowledge across technology and internationalisation. Patent indicators can serve as a measure of the output of R\&D, its productivity, structure and the development of a specific technology or industry. 

As indicators, patents have the following advantages over other measures i) a close link to invention, ii) they cover a broad range of technologies on which there are sometimes few other sources of data, iii) the contents of patent documents are a rich source of information, iv) patent data are readily available from patent office \cite{OECD2009}.
A popular composite indicator for measuring innovativeness at the national level is the Summary Innovation Index (SII) annually published in a report: The European Innovation Scoreboard. One of its indicators is the number of Patent Cooperation Treaty (PCT) applications per billion GDP  \cite{Hollanders2019}. 
Another well-know composite innovation indicator, which among other things contains quantifiable information about patents (patent families filed in two or more offices; Patent applications by origin; PCT international applications by origin), is the Global Innovation Index, published by Cornell University, INSEAD and the World Intellectual Property Organization \cite{Dutta}.

It is not only big organizations analysing patents. Independent scientific researchers also often utilize patents as an indicator of country innovation performance e.g.~\cite{Lee2020}. 
\cite{Kim2015} moves the analysis further by focusing on the different natures of national innovation systems in East Asia and Latin America. The key conclusion is that it is not scientific knowledge (academic articles) but technological knowledge (patents) that matters for economic growth. Furthermore, generating scientific knowledge does not automatically lead to the generation of technological knowledge. They noticed that technological knowledge is primarily determined by corporate research and development efforts, which used to be more lacking in Latin American countries, compared to East Asia.

Companies invest in patents because, new inventions through efforts on R\&D activities get protected by approval of patents rights \cite{Das2020}. 
The patent system is one of a suite of policy levers that has been used to attempt to bring the private returns captured by inventors closer to the social value of their inventions.  
Patents aim to allow inventors to recoup the fixed costs of their research investments by providing inventors with a temporary period of market power \cite{Williams2017}. 
A large amount of literature in economics, management, finance, law, and related fields has developed over the past few decades to investigate various aspects of the patent system. 

Having said that, it is worth noting that there is a contradictory role of patents in improving innovation found in the literature. 
On one hand, \cite{Haber2016} summarized as follows \textit{``… the weight of the evidence supports the claim of a positive causal relationship between the strength of patent rights and innovation''}. 
On the other hand, \cite{Boldrin2013} presents contrasting argument against patents \textit{``The case against patents can be summarized briefly: there is no empirical evidence that they serve to increase innovation and productivity…''}.
Investigations of the relationship between the changes in patent activity and the amount of expenditure on R\&D are not novel e.g.  \cite{Griliches1990}.
The conclusion is that the increase in the patent activity of large enterprises can be achieved by increasing expenditure on R\&D. 

\cite{Sierotowicz2015} evaluated the efficiency of R\&D expenditure (data obtained from EUROSTAT) from the patent activity (data obtained from EPO) in EU-28 countries for the period 1999–2013. Among the 28 EU counties, 98.07\% of assignee patents granted by the EPO in that period were given to entities belonging to the business enterprise sectors of the leading countries, which included: Germany, followed by France, United Kingdom, Italy, Sweden, Netherlands, Finland, Austria, Denmark and Spain, while the remaining 1.93\% of assignees belong to the remaining 18 countries of the EU.  
The author concluded that the increase in total intramural expenditure on R\&D activities in the business enterprise sectors of the 10 leading EU countries caused the increases in patenting activity of the sector in the long run. It was presented that Germany has the highest value of 0.26 of patents granted by enterprises per 1 million euros of the total intramural expenditure on R\&D in the business enterprise sector across the research period, while the lowest value is Spain, at 0.03.

Empirical evidence on the link between firms’ R\&D expenditure and patent registrations in Spain was provided by~\cite{Altuzarra2019}. 
A bidirectional causal relationship between R\&D and patents was evidenced by the Granger causality test in a panel of Spanish manufacturing firms for the period 1990–2013.
A broader study of \cite{Almeida2007} on whether patenting negatively impacts R\&D activity in a panel of 88 countries over an eight-year period (1996–2003) found mixed support for the negativity of patents on R\&D investment. Accumulated patents positively impact on R\&D intensity for the set of less developed countries whereas no statistically significant effect emerges in the case of higher developed converge clubs. When restricting the highest developed convergence club down to countries with a R\&D intensity above 3\%, the reversed negative causality arises, corroborating the asymmetric impact of patents on R\&D investment.

The review of existing studies covers different aspects of patents e.g. interlinkages among R\&D; patents and incomes in different countries or comparative analysis between countries. Also, multiple reports point to the issue of under-performance in many aspects by the EU-13 in comparison with the EU-15 Member States. The effects of EU accession has been studied by many authors, however, the particular aspect of innovations could be identified only in a few scientific publications.

\cite{gyen_2018} investigated the relationship between innovation and EU membership using panel data at the business level by using a difference-in-difference estimator considering access to the inner EU market as the treatment. The findings are that there is a significant percentage point decline in innovation efforts by firms in the new membership countries relative to the change for the control group firms.  
\cite{Makkonen2016} find that the most significant impact, in terms of co-publication intensities, of the EU enlargement, has been the high increase in the level of scientific collaboration that the EU-10 (new member states of the 2004 enlargement) and EU-2 (new member states of the 2007 enlargement) have among each other. Additionally, the collaboration between the new and old member states has been affected by the EU enlargement. In both papers, author use causal model for their analysis.

Similarly \cite{mtar2020causal} examined the causal relationship between innovation, financial development, and economic growth using panel VAR approach for 27 OECD countries over the period 2001–2016. Among others, authors conclude that the relationship between innovation and economic growth is complex and country-specific characteristics can play an important role in fostering innovation and productivity. The paper concludes that governments can play an important role in developing a legislative framework favoring the development of innovation financing through the patent guarantee deposit.

Many additional factors have been listed in the literature, which could affect innovation.
Having said that, multiple findings and applications show that despite the complicated nature of innovation, patents are one of the key measurements of countries innovation performance. To the best of the author’s knowledge, there are no other papers aiming at quantifying the causal effect of EU accession on patents performance for specific EU countries.

Causal modeling techniques are not novel and are widely used to test causal claims both within economics and in many other areas of social sciences e.g \cite{Woodward1995,morgan_winship_2014}. The aforementioned papers use this method frequently. Quasi-experimental methods, such as difference-in-difference, regression discontinuity and other related methods, have had the effect of overshadowing the role of economic theory in the specification of a model.

Aforementioned Granger causality is based on autoregressive (AR) processes applicable to problems of model identification, while the transfer entropy method is an information-theoretic approach that does not need assumptions on the structure of the process. It is based on the concept of Shannon-entropy and is suitable for linear and nonlinear relations. Its key assumption is that the sampled data should follow a well-defined probability distribution. Both approaches are however essentially descriptive, as they are not based on a structural modeling of the data generating process. This approach could be found e.g.\cite{Altuzarra2019}. 
However this method could have some limitation: Granger-causality may not be sufficient in practice for counterfactual control as identified by~\cite{HOOVER201289}.

The difference-in-difference (DiD) approach is a quantitative research design for estimating causal relationships in quasi-experimental settings. It is popular, for example, in empirical economics as well as other social sciences and commonly applied when estimating the effects of certain policy interventions or institutional changes that do not affect everybody at the same time. This approach could be found in e.g  \cite{Makkonen2016}.
However, DiD is limited, for example:i) DiD is traditionally based on a static regression model that assumes i.i.d. data despite the temporal aspect of the data. When fit to serially correlated data, static models yield overoptimistic inferences with too narrow uncertainty intervals; 
ii) most DiD analyses only consider the point in time when the intervention happened, and no evolution of the effect can be inferred; 
iii) Synthetic control construction can be restricted in the case of time-series analysis.

All the issues of DiD are fixed by Google's (\cite{causalimpact}) proposal called \emph{CausalImpact}.
The method generalises DiD to the time-series setting by using Bayesian structural time-series models to construct a counterfactual effect estimator.
Multiple advantages of the methods described in section~\ref{sec:materialsmethods} make the method particularly suited to the needs of this paper.

\section{Materials and methods}
\label{sec:materialsmethods}

In this section we present formal methods of assessment if EU membership change patents performance for new members (EU-13).
The method from~\cite{causalimpact} was used in the paper as a primary causal analysis method. All causal impact calculations were performed with Google’s \emph{CausalImpact} R package.

Patent information crucial for the analysis can be obtained from different sources.
Starting from publicly available databases like  EPO, OECD or Google Patents to the more advanced commercial ones like Derwent, PATSTAT or Global Patent Index. 
The analysis reported in this paper is based on a dataset from the Organisation for Economic Co-operation and Development (OECD).

\subsection{Dataset}

We used the OECD database~\cite{oecdstat}
where patents were counted according to the inventor’s country of residence. 
When a patent was invented by several inventors from different countries the respective contributions of each country was accounted for in order to eliminate multiple counting of such patents (hence fractional counts existed in the dataset). 
The research concerns the number of yearly patents 
in 28 EU countries: CY- Cyprus, MT-	Malta, LV- Latvia, LT- Lithuania, EE- Estonia, HR- Croatia, BG- Bulgaria, SK- Slovak Republic, RO- Romania, LU- Luxembourg, SI- Slovenia, PT- Portugal, EL- Greece, CZ- Czech Republic, HU- Hungary, PL- Poland, IE- Ireland, DK- Denmark, ES- Spain, BE- Belgium, FI- Finland, AT- Austria, SE- Sweden, NL- Netherlands, IT- Italy, UK- United Kingdom, FR- France, DE- Germany. 
The analysis is carried out on  patents collected over the years 1985–2017.
A whole group of countries, namely the so-called New Member States that joined the European Union in 2004 (CZ, EE, HU, LV, LT, PL, SK, SI, MT and CY) 2007 (BG, RO) and 2013 (HR) will be collectively referred to as the EU-13 with EU-10, EU-2, EU-1 denoting the subgroups respectively. Old Member States countries (AT, BE, DE, DK, EL, ES, FI, FR, IE, IT, LU, NL, PT, SE, UK) we will refer to as the EU-15. For each EU-13 country the year of EU accession divides the patent’s dataset for that particular country into two time-series refereed to as  \emph{before accession} and \emph{after accession}.

\subsection{Causal impact}
Causal modelling enables reasoning about the cause and the effect in contrast to correlation models where only the association can be reasoned about~\cite{pearl2009causality}.
The causal impact of a treatment is the difference between the observed value of the response and the value that would have been obtained under the alternative treatment, that is, the effect of treatment on the treated. 
In this paper the response variable is a time series $y_t$, so the causal effect of interest is the difference between the observed series and the series that would have been observed had the intervention not taken place.
In the research reported here, the intervention is the EU accession and the response variables represent yearly patents assigned to the country. As the global political environment is not an isolated experimental environment it is not possible to measure the counterfactual response. Thus we applied a method proposed by~\cite{causalimpact} that uses time-series prediction to estimate the counterfactuals. 

A time-series model fitted to the observations from before intervention can predict what would have happened if the country had not accessed the EU. The uncertainty of the prediction is handled by using Bayesian structural time-series models. This allows estimation of the statistical significance of the causal impact. Since prediction intervals often grow fast as the number of predicted steps increases, a simple time series model is insufficient to conduct a meaningful analysis. Time-series observations from the countries not affected by the intervention (yet correlated with the analyzed time series) are crucial for the model to capture global trends. In this paper we used patent time-series for the EU-15 to reduce prediction uncertainty for the new EU countries. Such a decision is motivated by research of ~\cite{Makkonen2016}, where EU-15 formed a control group in a DiD method. The next section describes mathematical details of the time-series model used for prediction.

\subsection{Linear Gaussian State Space Model}

Linear Gaussian State Space Model (LG-SSM) is a general family of time series models. 
\cite{murphy2012machine} defines it as the following linear dynamical system:
\begin{align}\label{eq:state_dynamics}
	\bm z_{t} &= \bm F_t\bm z_{t-1}+\bm R_t\bm\varepsilon_t, \quad \bm\varepsilon_t\sim \mathcal{N}(\bm 0,\bm Q_t),\\
	\label{eq:state_dynamics2}
	y_t&=H_t^T\bm z_t+\delta_t, \quad \delta_t \sim \mathcal{N}(0,\sigma_y^2),
\end{align}
where $\bm z_0\sim \mathcal{N}(\bm b_0,\bm Q_0)$ is the $n$-dimensional initial hidden state and $ x_t\in\mathbb{R}$, $t\ge 0$, is the observed time series. Throughout, $\mathcal{N}(\cdot,\cdot)$ denotes a multivariate Gaussian distribution and the noises $(\bm\varepsilon_t,\delta_t)$ are independent across time $t$.
The state dynamics are parameterized by the  transition matrix $\bm F_t\in\mathbb{R}^{n\times n}$, control matrix $\bm R_t\in\mathbb{R}^{n\times q}$, and covariance matrix $\bm Q\in\mathbb{R}^{q\times q}$.
The observations $x_t$ are noisy linear projections of the states $\bm z_t$ and further parameterized by the observation vector $H\in\mathbb{R}^{ n}$ and the observation noise variance $\sigma_y^2$.
Many classical time series models can be represented as an LG-SSM, this makes it a popular choice for time series forecasting~\cite{harvey1990forecasting}.
In this paper we focus on two main components: local level and linear regression.

Mathematically a local level model is  defined as LG-SSM with a real hidden state, namely, a current level $l_t$.  
It evolves as:
\begin{align}
	y_t^l &= l_t+\epsilon_t^{y}\quad \epsilon_t^{y}\sim\mathcal{N}(0, \sigma_y^2)\\
	l_t &= l_{t-1}+\epsilon_t^{l}\label{eq:level}\quad \epsilon_t^{l}\sim\mathcal{N}(0, \sigma_l^2)
\end{align}
where $\epsilon_t^{y}$ and $\epsilon_t^{l}$ are independent.
The future depends only on the past observations of the time-series.

Often time series can be explained by another time series and in addition to its past.
In LG-SSM it is possible to use external observations (time-series) $\bm x_t$ in the form of linear regression.
A \emph{static linear regression} is obtained by setting  $H_t=\bm\beta^T \bm x_t$ and $z_t^r=1$, where
$\bm x_t$ is a vector of covariates (in our case the EU-15 with additional column of ones) and $\bm\beta$ is vector of regression coefficients.
Since the sum of LG-SSMs is also an LG-SSM the simple components (local level and linear regression) can be summed to form structural time series model:~\cite{murphy2012machine}:
\begin{align}
    y_t &= l_t+\bm \beta^T \bm x_t + \epsilon_t^{y},\label{eq:model} 
\end{align}
Where $l_t$ is the local level~\eqref{eq:level}.
In terms of the general representation from \eqref{eq:state_dynamics} we obtain a block-diagonal transition and control matrices and concatenated observation vectors performing the summation of local level and regression components in ~\eqref{eq:model}.
Given the model parameters, the state $\bm z_T$ at time of intervention and the contemporaneous predictor variables after intervention $\bm x_{t\geq T}$, the system dynamics~\eqref{eq:state_dynamics}--\eqref{eq:state_dynamics2} enable forecasting for $y_{t\geq T}$.
Since the model is linear and Gaussian, the forecast has a normal distribution. If the patent data were just counts, a hierarchical model with Poisson distribution for the response variable would be required. Having said that we emphasize that the patents in our dataset are fractional so we use vanilla LG-SSM.

\subsection{Bayesian inference}

In order to fit LG-SSM to the patent data of interest, we follow a Bayesian approach that captures uncertainty via a posterior distribution as deeply described by~\cite{greenberg_2012,bayes:nature}.   
For Bayesian inference in LG-SSM a joint prior distribution is to be specified for the initial states and the unknown model parameters.  

The prior for the local level diffusion scale ($\sigma_l$) is inverse Gamma interpreted as a scaled inverse-$\chi^2$ distribution (as described by~\cite{gelman2013bayesian}) with standard deviation $0.1\hat\sigma_y$ and 32 degrees of freedom, where $\hat\sigma_y$ is standard deviation of the time series. 
The prior is truncated to maximum value at $\hat\sigma_y$.
Initial state prior is Normal with mean equal to the first  observation $y_0$ of the time series and scale equal  to $\hat\sigma_y$.
For the regression we used Zellner's g-prior with spike-and-slab.
The prior for $\bm \beta$ and $\sigma_y$ is commonly expressed in terms of expected model size (in our case equal to 3), expected $R^2$ (0.8) and degrees of freedom (50). 
Spike-and-slab prior enforces sparsity of $\bm \beta$, i.e. most of the values will be exactly 0.
Such a prior acts as a Bayesian model selection algorithm and allows the addition of multiple covariates in the model without overfitting. 
Only the most important covariates get nonzero coefficient thus they are present in the model.  All the other variables are multiplied by zero so they are not part of the model.
The hyperparameters for these distributions are set  using heuristics from \emph{CausalImpact} package that  aim to provide default 
prior distributions.
In order to sample from the posterior distribution we apply Monte Carlo Markov Chain (MCMC)~\cite{murphy2012machine}.

\section{Results}
\label{sec:results}

\begin{figure}
 \caption{Number of patents before and after accession with the percentage of increase for a given country. The diagonal line marks zero change due to accession.}
 \label{fig:beforeafter}
    {
    \centering
      \includegraphics[width=2\mywidth]{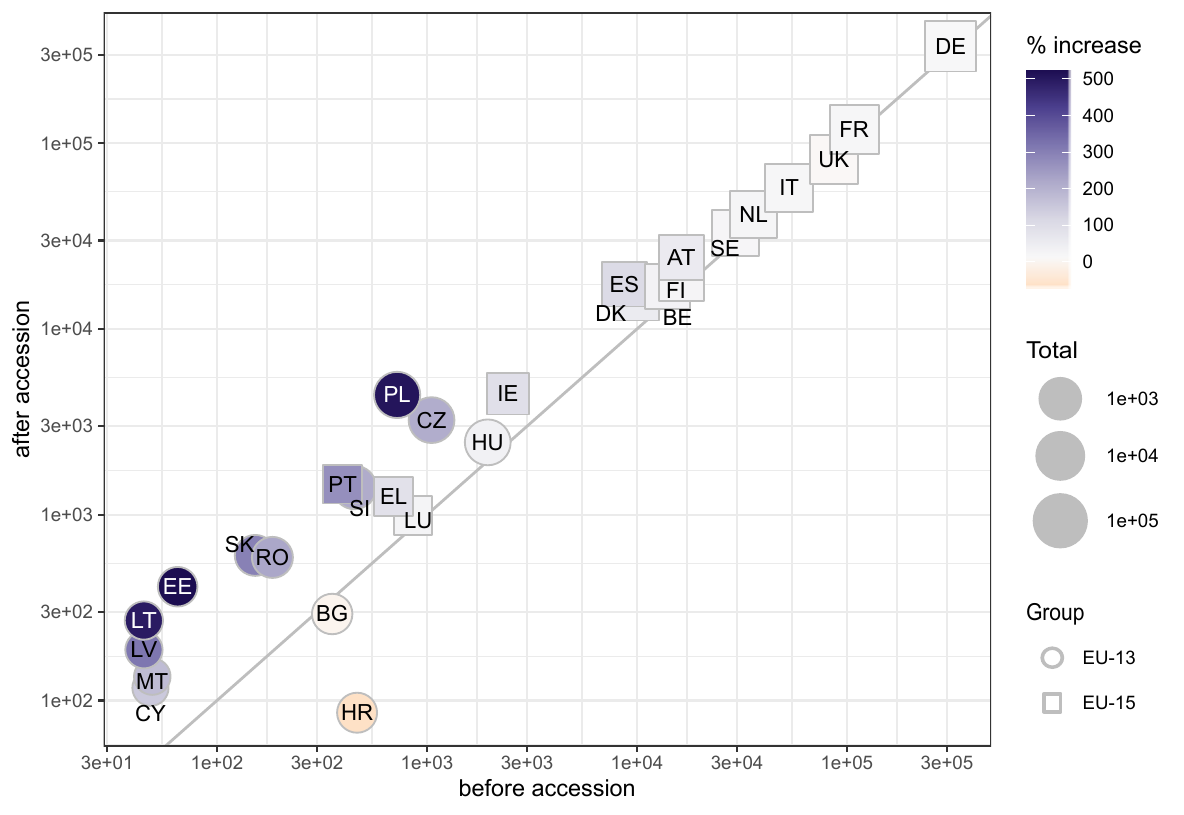}
    }
\end{figure}

As Figure \ref{fig:beforeafter} shows, the number of patents (as a sum in years 1985-2017) among the EU member states varies greatly between countries. 
For better visualization we present EU-13 and EU-15 in logarithmic scale (maximum number of patents for country from EU-15 (Germany) was above 645,912 while in EU-13 (Poland) above 5,000). The greatest number of patents from EU-15 was assigned to Germany, followed by France, United Kingdom, Italy and the Netherlands. The highest number of patents from the EU-13 was assigned to Poland, Hungary and Czech Republic. Also, the fraction of countries from the EU-13 (when counting the total number of patents) was negligible compared to the EU-15. In total, 98.66\% (1,462,149.9) of patents during the analyzed period were allocated in fifteen EU-15 countries while in the EU-13 countries only 1.34\% (19,783.4) were allocated. This difference represents a huge dominance of the EU-15 in a number of patents. Even accounting for the fact that countries from the EU-13 account for approximately 20\% of the EU population, it is still a negligible amount. Interestingly from a visual perspective (considering two periods before and after accession), we could observe that the fraction of patents after accession is higher (points above diagonal line in Figure~\ref{fig:beforeafter}) in the EU-13, while for the EU-15 there was no big change (corresponding points lay nearly diagonal line).

 Simple analysis of total number of patents before accession and after accession in terms of percentage increase reveals the highest increase of patents occurred in EU-13 countries suggesting that EU accession affected patents performance in those countries see Figure \ref{fig:beforeafter}.
The decrease was noticed (points lying below the diagonal line) for Bulgaria (-17) and United Kingdom (-5.4).
The result for Croatia is incomparable because of an uneven  period 1985- 2012 versus 2013-2017.
The lowest increase was observed for Germany (6.8), Italy (8.5) and France (9.8).  
Whereas the number of patents increased over the two periods especially in Estonia (+530), Poland (+512) and Lithuania (+500). 
In general the highest increase of patents performance during the period considered was in the EU-13 and the lowest in EU-15 countries. However, from this visual examination, it cannot be inferred with statistical precision if the growth can be explained by the overall increasing trend over the last decades or whether joining the EU caused the changes.

Thus, we apply a formal statistical causal modeling approach which allows us to analyze whether EU membership has led to a statistically significant growth in the number of patents in EU-13 countries compared to the counterfactual scenario that each country had not accessioned the EU. We rely on the patent dynamics before accession and the number of patents in the EU-15 that received no treatment, for obtaining accurate predictions for the counterfactual scenario. Summaries of the analysis obtained for 20,000 MCMC samples is collected in Table~\ref{tab:casumm}.

The rows with \emph{Stat} Average (Avg.) refer to the average patents (over the time period) during the post-intervention period.
The rows with \emph{Stat} Cumulative (Cum.) represent the sum of the individual yearly observations.
The interpretation of our results for a given country e.g. Poland (which could be generalized for the rest of the countries- accordingly to their values) is as follows. 
In the post-intervention period, the response variable had an average value of approx. 312.2 as reported in column \emph{Actual}. 
By contrast, in the absence of an intervention, we would have expected an average response of 101.1 (see: column \emph{Val}). 
The 95\% interval of this counterfactual prediction is [22.4 - 228.6 ] (see: columns \emph{lower} and \emph{upper} in \emph{Prediction} group). 
Subtracting this prediction from the observed response yields an estimate of the causal effect that the intervention had on the response variable.  
The same interpretation applies to entries in a row (\emph{Cum}).
In relative terms, the response variable showed an increase of +212\%
(see column \emph{Val} in \emph{Relative} group). 
The 95\% interval of this percentage is [+84\%, +290\%] (respectively columns \emph{lower} and \emph{upper} in \emph{Relative} group ) means that the positive effect observed during the intervention period is statistically significant.
Note that this increase is the causal effect relative to a predicted counterfactual value, not to be confused with the percentage from Figure~\ref{fig:beforeafter}.

Statistical significance of the analysis can be assessed from the last column \emph{p}. 
The effect observed during the intervention period is statistically significant and unlikely to be due to random fluctuations if the probability of obtaining this effect by chance is very small (Bayesian one-sided tail-area probability \emph{p} ).
For countries where $p<0.05$ (posterior prob. of a causal effect equal $1-p>=95\%$) the causal effect can be considered statistically significant, otherwise the effect may be spurious and would generally not be considered statistically significant.
A significant causal impact on patents performance after accession can be observed in Romania, Estonia, Poland, Czech Republic, Croatia and Lithuania. However, as results from analysis show for Croatia and Lithuania, joining the EU caused a negative impact for patents performance. For the remaining countries: Hungary, Latvia, Cyprus, Slovakia, Slovenia, Bulgaria and Malta joining the EU did not statistically significantly affect patents performance. An interesting result is observed for Latvia where the percentage change increase (see Figure \ref{fig:beforeafter}) could misleadingly suggest significant influence after accession (above 300\% increase after accession) however, model prediction classified this increase as statistically not significant. It means that this increase came from trends before accession, so such growth is not influenced by EU Membership.
\begin{table}[h!]
    \centering
    \caption{Summaries of causal impact analysis}
    \captionsetup[table]{labelformat=empty,skip=1pt}
\begin{tabular}{l|lrrrrrrrr}
\toprule
\multicolumn{3}{c}{} & \multicolumn{3}{c}{\textbf{Prediction$\pm$95\% CI}}& \multicolumn{3}{c}{\textbf{Relative $\pm$95\% CI}}&\\
\cmidrule(r){4-6}\cmidrule(r){7-9}
 & \textbf{Stat} & \textbf{Actual} & \textbf{Val} & \textbf{lower} & \textbf{upper} & \textbf{Val} & \textbf{lower} & \textbf{upper} & \textbf{p} \\
\midrule
RO & Avg & $54.0$ & $29.3$ & $20.6$ & $38.5$ & $85\%$ & $53\%$ & $114\%$ & $0.000$ \\ 
RO & Cum & $540.0$ & $292.6$ & $205.6$ & $385.2$ & $85\%$ & $53\%$ & $114\%$ & $0.000$ \\ 
EE & Avg & $30.4$ & $13.9$ & $7.1$ & $23.5$ & $118\%$ & $50\%$ & $168\%$ & $0.000$ \\ 
EE & Cum & $394.9$ & $180.9$ & $91.7$ & $305.1$ & $118\%$ & $50\%$ & $168\%$ & $0.000$ \\ 
PL & Avg & $312.2$ & $100.1$ & $22.4$ & $228.6$ & $212\%$ & $84\%$ & $290\%$ & $0.001$ \\ 
PL & Cum & $4,058.5$ & $1,301.0$ & $291.7$ & $2,971.3$ & $212\%$ & $84\%$ & $290\%$ & $0.001$ \\ 
CZ & Avg & $229.6$ & $126.3$ & $20.5$ & $223.1$ & $82\%$ & $5\%$ & $166\%$ & $0.018$ \\ 
CZ & Cum & $2,984.7$ & $1,641.5$ & $266.9$ & $2,900.6$ & $82\%$ & $5\%$ & $166\%$ & $0.018$ \\ 
HR & Avg & $19.2$ & $27.9$ & $18.0$ & $36.7$ & $-31\%$ & $-63\%$ & $5\%$ & $0.040$ \\ 
HR & Cum & $77.0$ & $111.8$ & $71.8$ & $147.0$ & $-31\%$ & $-63\%$ & $5\%$ & $0.040$ \\ 
LT & Avg & $20.1$ & $33.3$ & $17.3$ & $46.2$ & $-40\%$ & $-78\%$ & $9\%$ & $0.047$ \\ 
LT & Cum & $261.7$ & $432.8$ & $224.7$ & $601.2$ & $-40\%$ & $-78\%$ & $9\%$ & $0.047$ \\ 
\midrule
HU & Avg & $183.8$ & $96.1$ & $-105.0$ & $224.1$ & $91\%$ & $-42\%$ & $300\%$ & $0.071$ \\ 
HU & Cum & $2,388.9$ & $1,249.8$ & $-1,365.3$ & $2,913.9$ & $91\%$ & $-42\%$ & $300\%$ & $0.071$ \\ 
LV & Avg & $14.1$ & $9.5$ & $4.3$ & $18.2$ & $48\%$ & $-43\%$ & $102\%$ & $0.115$ \\ 
LV & Cum & $183.3$ & $123.8$ & $56.5$ & $236.7$ & $48\%$ & $-43\%$ & $102\%$ & $0.115$ \\ 
CY & Avg & $8.8$ & $6.3$ & $1.7$ & $11.0$ & $40\%$ & $-35\%$ & $113\%$ & $0.133$ \\ 
CY & Cum & $114.5$ & $81.8$ & $21.8$ & $142.9$ & $40\%$ & $-35\%$ & $113\%$ & $0.133$ \\ 
SK & Avg & $43.3$ & $33.3$ & $15.7$ & $55.5$ & $30\%$ & $-37\%$ & $83\%$ & $0.217$ \\ 
SK & Cum & $562.9$ & $433.0$ & $204.7$ & $721.1$ & $30\%$ & $-37\%$ & $83\%$ & $0.217$ \\ 
SI & Avg & $101.1$ & $93.9$ & $53.8$ & $139.4$ & $8\%$ & $-41\%$ & $50\%$ & $0.353$ \\ 
SI & Cum & $1,314.5$ & $1,221.0$ & $699.9$ & $1,812.2$ & $8\%$ & $-41\%$ & $50\%$ & $0.353$ \\ 
BG & Avg & $27.4$ & $24.8$ & $-7.0$ & $72.1$ & $11\%$ & $-181\%$ & $139\%$ & $0.389$ \\ 
BG & Cum & $273.7$ & $247.7$ & $-70.3$ & $721.2$ & $11\%$ & $-181\%$ & $139\%$ & $0.389$ \\ 
MT & Avg & $9.7$ & $9.6$ & $4.2$ & $16.4$ & $1\%$ & $-70\%$ & $57\%$ & $0.445$ \\ 
MT & Cum & $125.9$ & $124.9$ & $54.0$ & $213.2$ & $1\%$ & $-70\%$ & $57\%$ & $0.445$ \\ 
\bottomrule
\end{tabular}

    \label{tab:casumm}
\end{table}

\begin{figure}
 \caption{Significant causal impact }\label{fig:sigimpact}
 \subfloat[Romania]{\includegraphics[width=\mywidth]{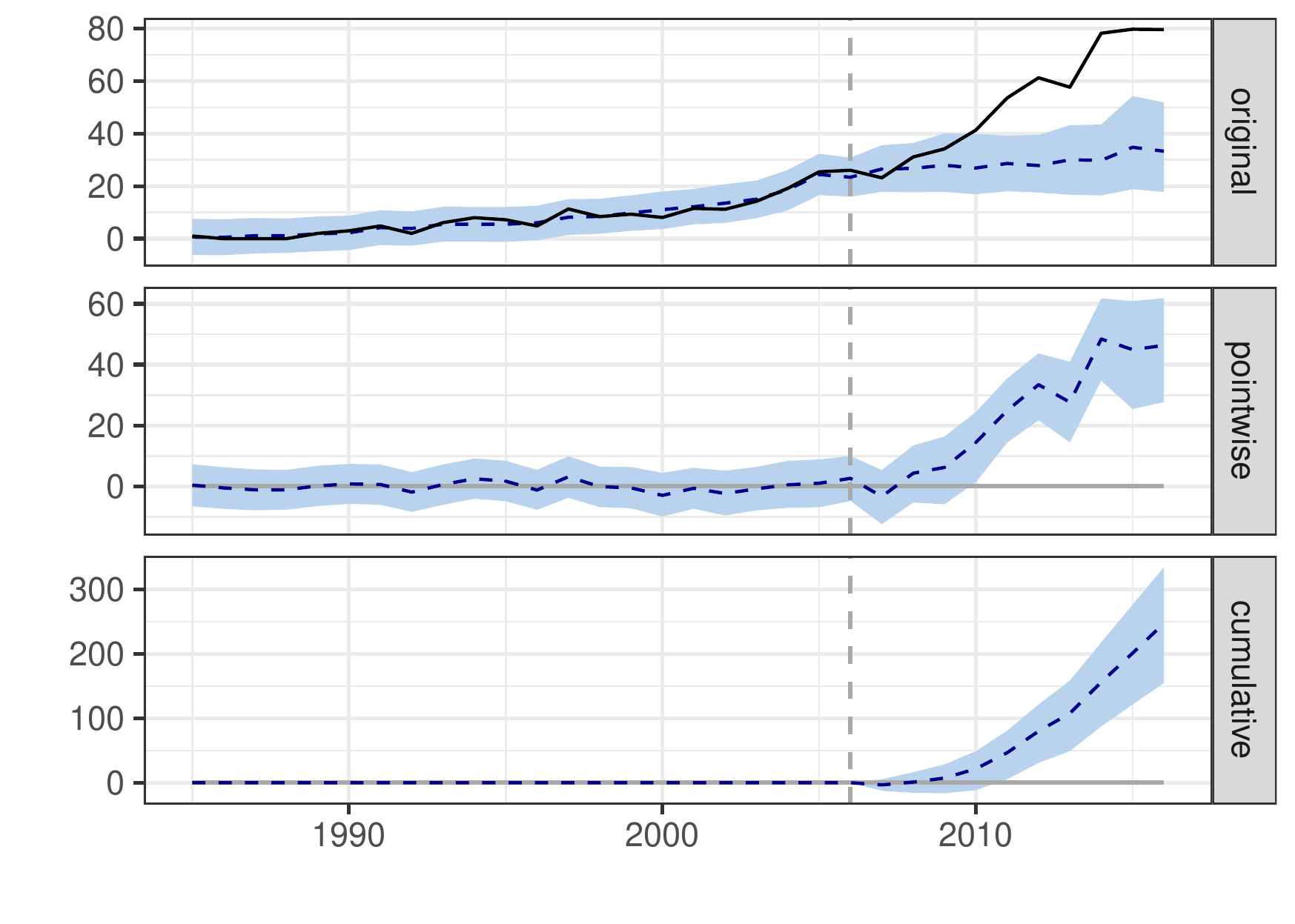}  }
 \hfill
 \subfloat[Estonia]{\includegraphics[width=\mywidth]{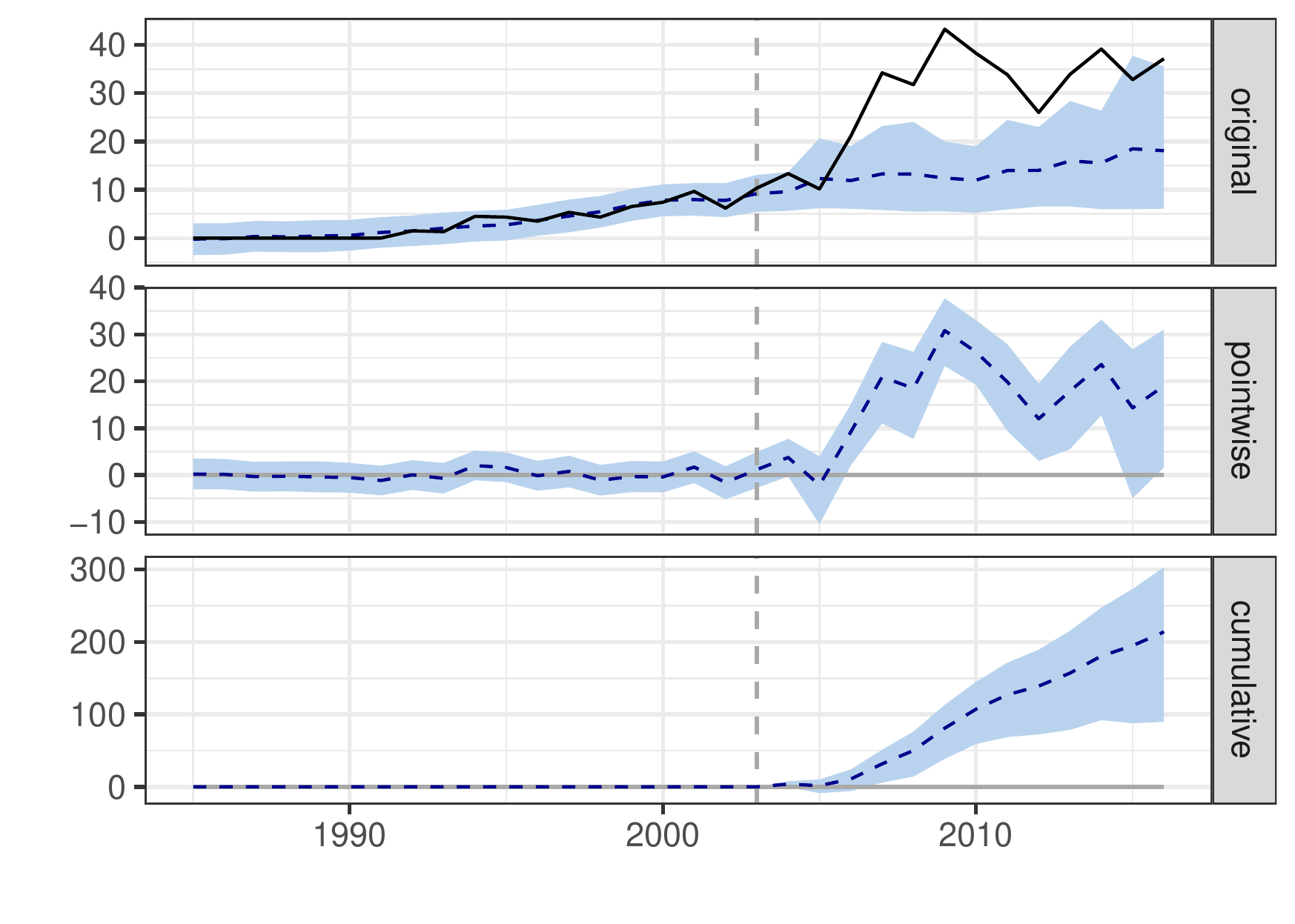}  }\\
 \subfloat[Poland]{\includegraphics[width=\mywidth]{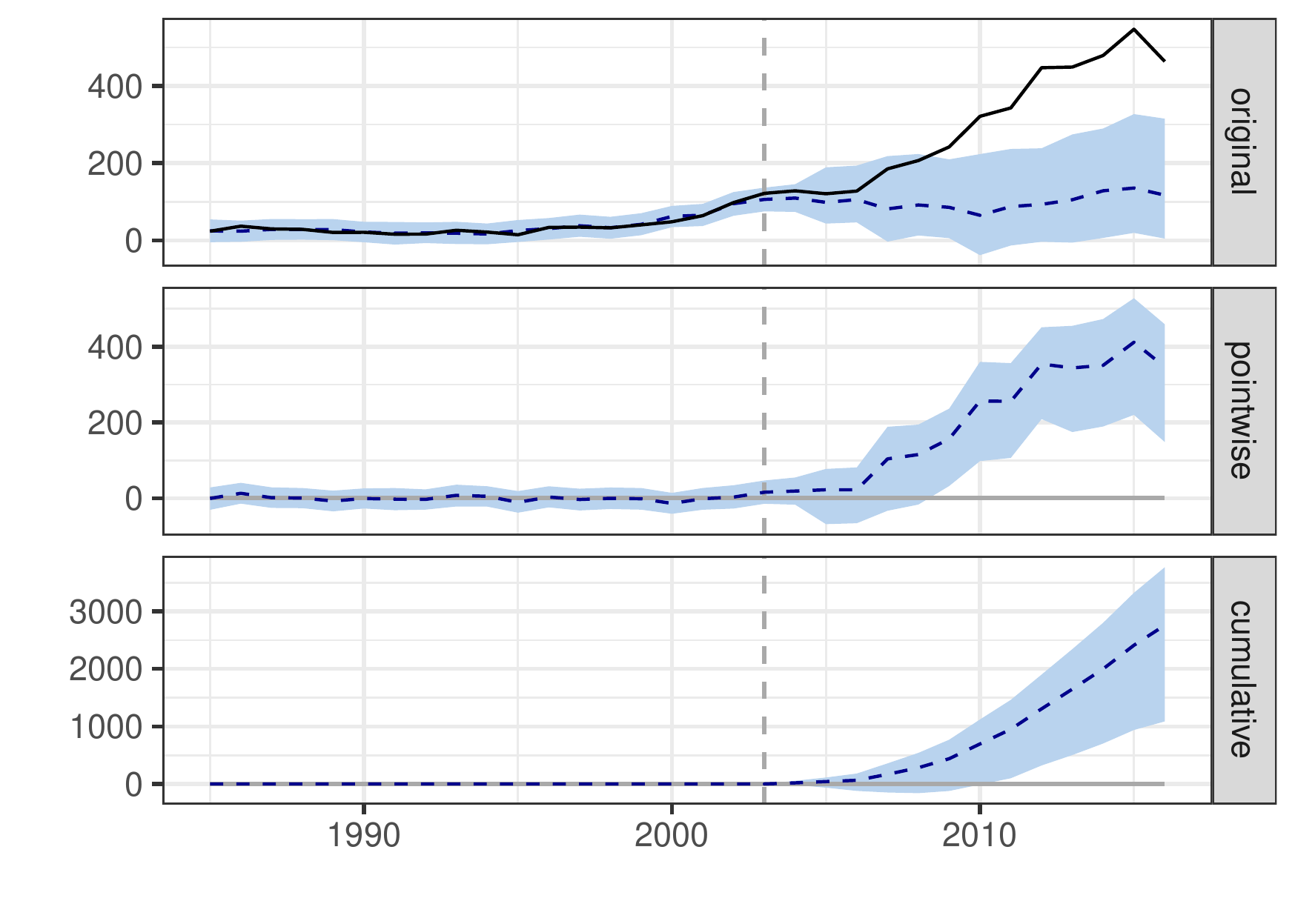}  }
 \hfill
 \subfloat[Czech Republic]{\includegraphics[width=\mywidth]{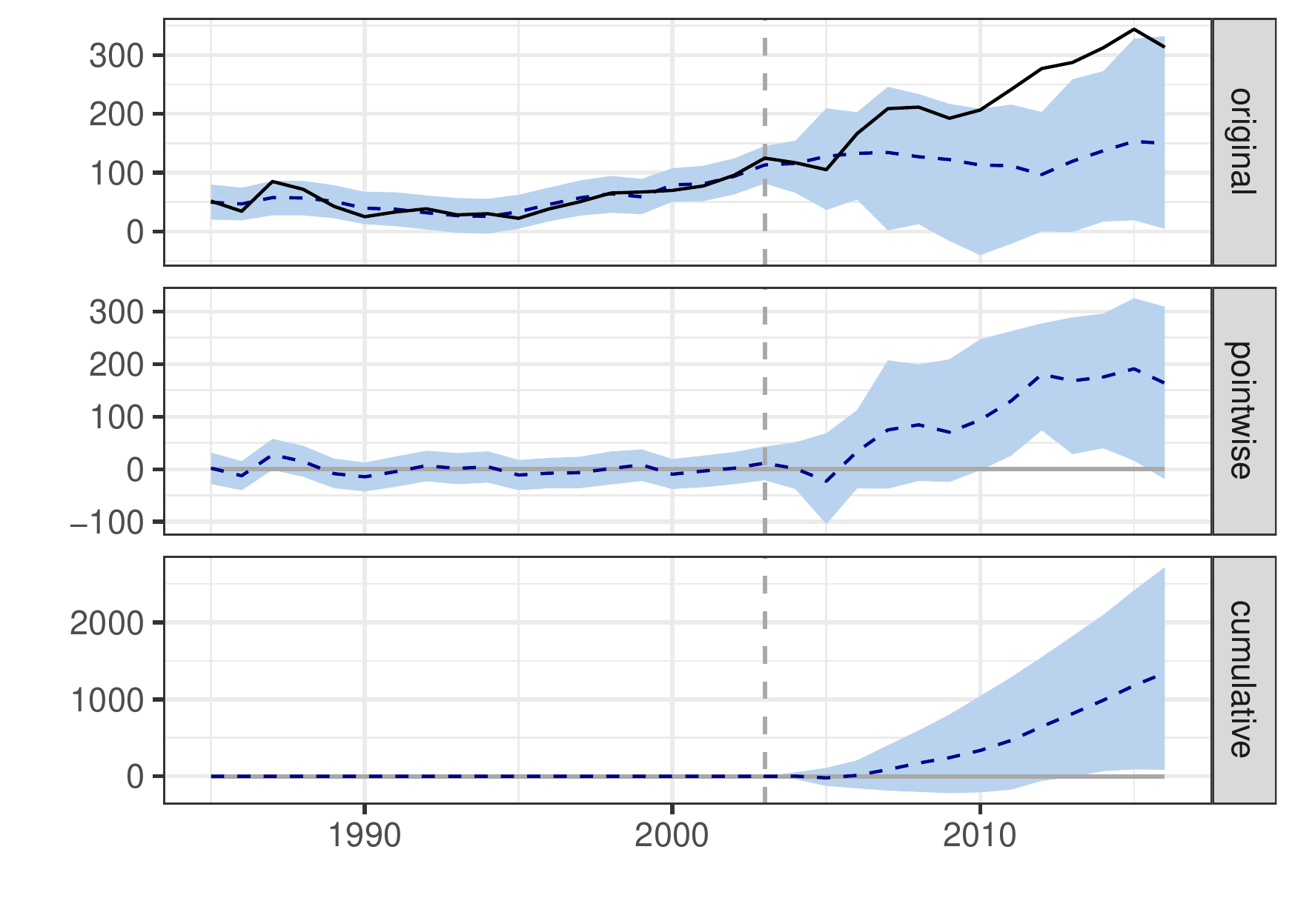}  }\\
 \subfloat[Croatia]{\includegraphics[width=\mywidth]{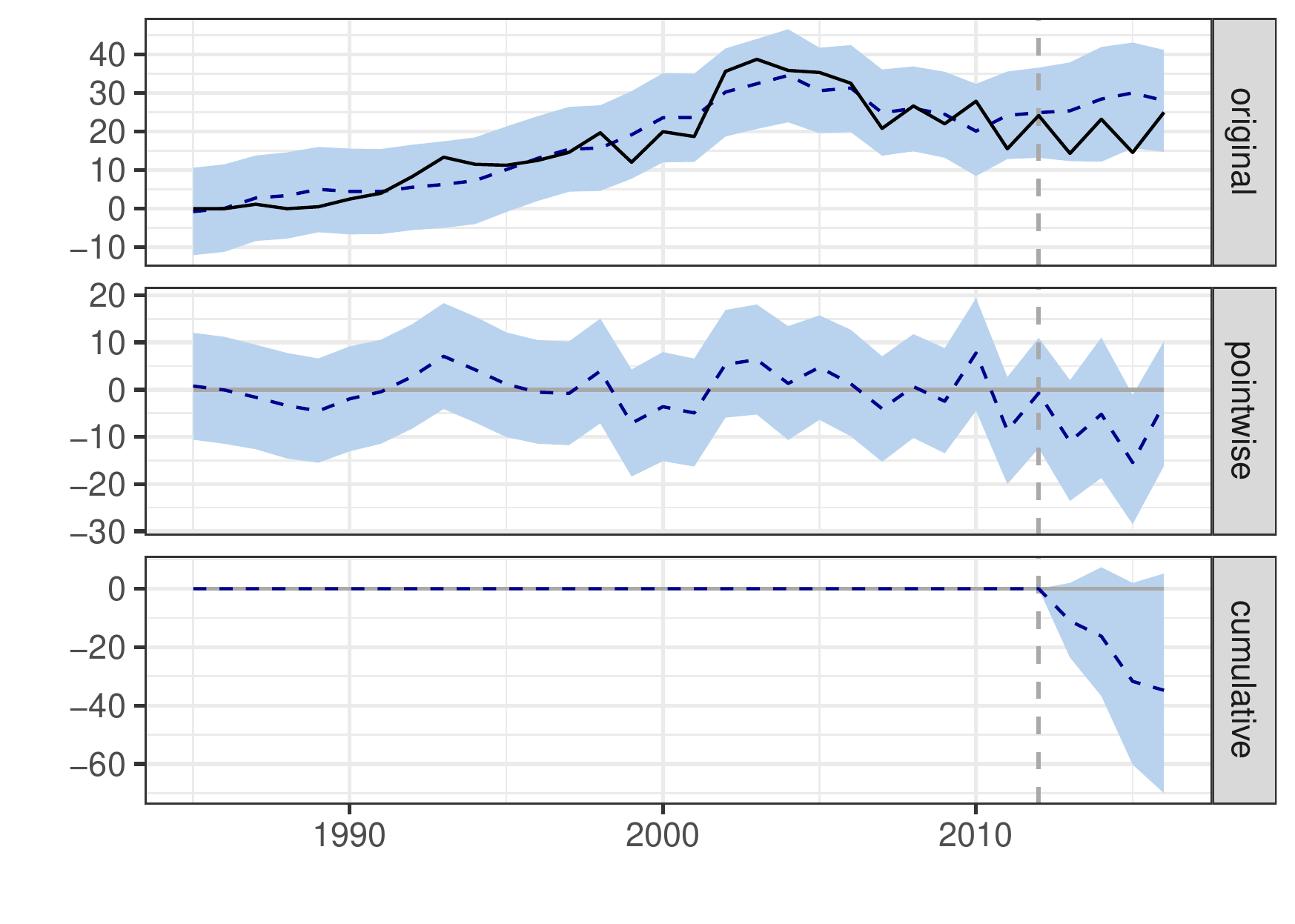}  }
 \hfill
  \subfloat[Lithuania]{\includegraphics[width=\mywidth]{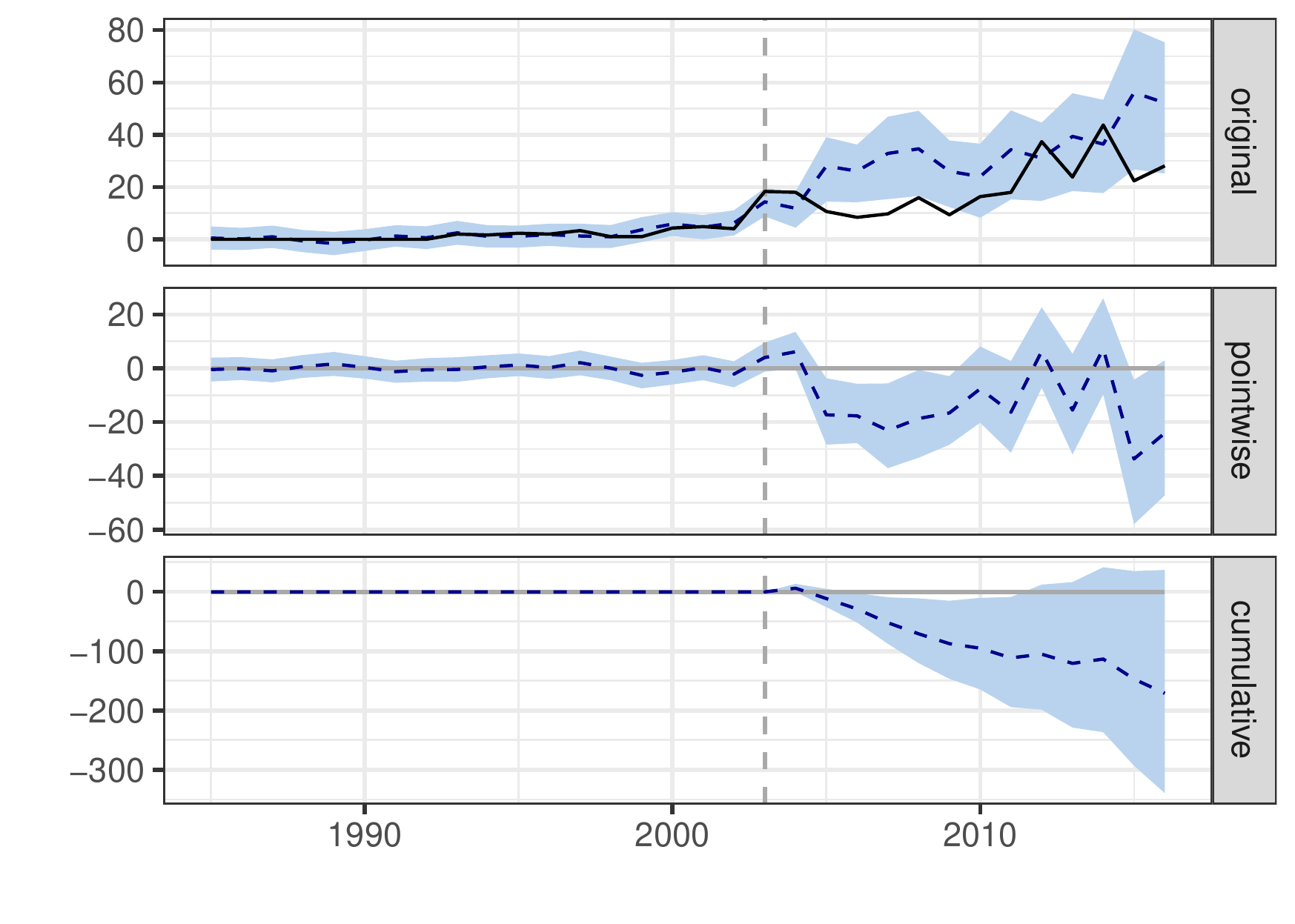}  }

    \source{own calculations}
\end{figure}

A detailed time evolution of the inferred causal impact can be observed  in Figures \ref{fig:sigimpact} and \ref{fig:insigimpact} for both groups of countries: with respectively significant  and insignificant  effect of EU accession on patents performance. 
Each plot consist of three separate charts:
\begin{description}
\item[original)] The observed time series (solid line) and fitted model’s forecast counterfactuals (dashed line) with 95\% credible interval (shaded area).
\item[pointwise)] The point-wise causal effect with 95\% credible interval, as estimated by the model. This is the difference between the observed outcome and the predicted outcome.
\item[cumulative)] The cumulative effect with 95\% credible interval -- the total number of patent due to EU accession.
\end{description}
The vertical line represents EU accession year (intervention) for the particular country (2004/2007/2013).
Further information that can be obtained from the plots include:
({\bf original}) the detailed time distribution (trajectory) of EU-13 countries.
({\bf pointwise}) The difference between observed data and counterfactual predictions is the inferred causal impact of the intervention. A key characteristic of the inferred impact series is the progressive widening of the posterior intervals (shaded area). This effect emerges naturally from the model structure that predictions should become increasingly uncertain as we look further and further into the future.
({\bf cumulative}) Another way of visualizing posterior inferences is by means of a cumulative impact plot. It shows, for each year, the summed effect up to that year. If the 95\% credible interval of the cumulative impact crosses the zero-line after the intervention, then at that point we would no longer declare a significant overall effect.

The same analysis performed for combined EU-13 countries shows significant positive impact of accession patent performance. All EU-13 countries in total have produced on average over 7,000 patents that can be attributed to the accession. However, surprisingly 7 of the EU-13 countries have not significantly benefited from the EU accession.

An interesting by-product of the causal analysis reported in this paper is a discovery of international relations in patent time series. 
Since coefficients in the linear regression are sparse due to spike-and-slab prior, only the most important covariates get nonzero coefficient.
In the model, the covariates for each of the EU-13 country are the patents for the EU-15 countries.
Let us consider two countries $c\in \text{EU-15}$ and $c'\in\text{EU-13}$.
If the coefficient $\beta_c$ is zero, it means that patent activity of  country $c$ does not explain the trend observed for country $c'$.
Contrary, a nonzero value of $\beta_c$ indicates that both countries have a similar dynamic of patents.
The similarity may originate from different socio-political interactions whose analysis is far beyond the scope of this paper. 
We only point out the existence of those pan-European interactions
\begin{figure}
    \caption{Fraction of nonzero regression coefficients }\label{fig:frac}
    {\centering
    \includegraphics[width=2\mywidth]{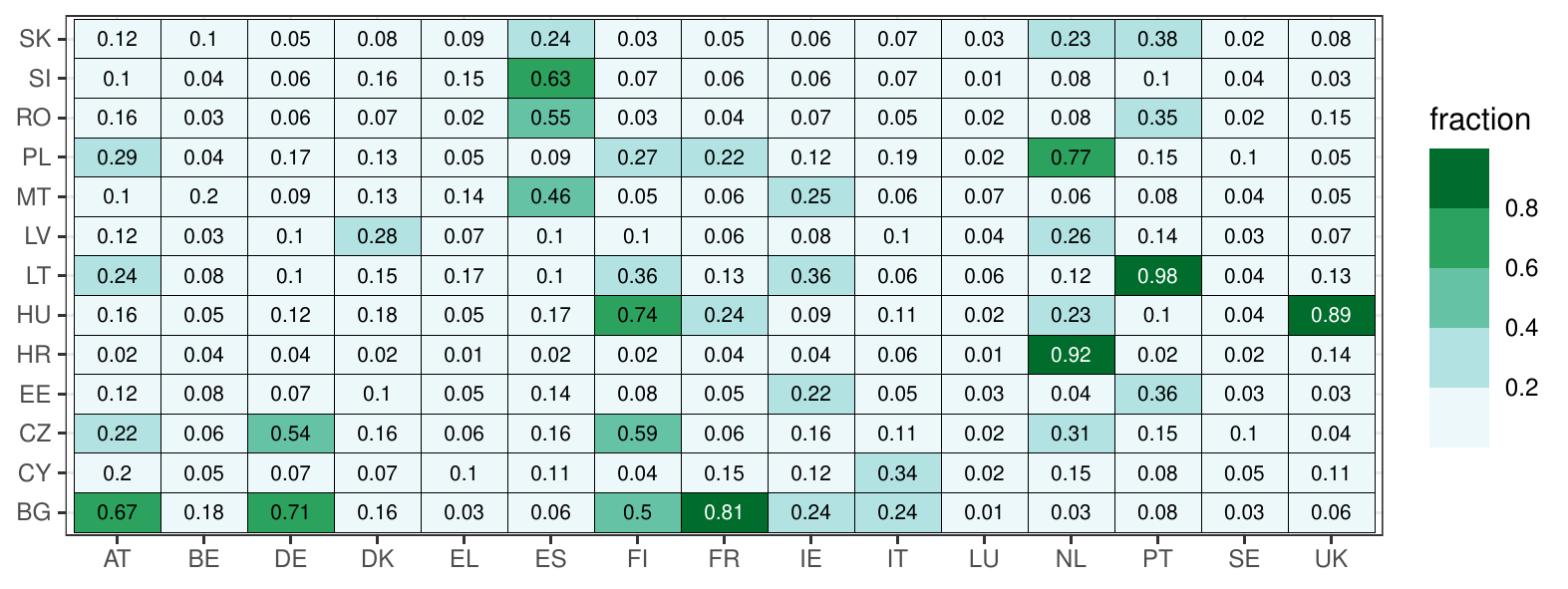}
    }
    \source{own calculations}
\end{figure}
In Figure~\ref{fig:frac} we visualized what fraction of 20,000 MCMC samples resulted in nonzero coefficient (estimate of slab probability). The $x$ axis represents covariates ($c$), while $y$ axis corresponds to the new EU members ($c'$).
Interestingly Luxembourg and Sweden participate very weakly in the model, while countries like Italy or Germany are frequently selected by the Bayesian model selection.

On the other hand, Hungary's patents are mostly explained solely by the time series for the Netherlands.
Similarly, the patent activity of Romania is mostly explained by data from Spain.
A deeper analysis of such relations could be helpful for policymakers in designing targeted research funds aimed at equalizing cooperation among the EU.

\begin{figure}
 \caption{Insignificant causal impact }\label{fig:insigimpact}
 \subfloat[Hungary]{\includegraphics[width=\mywidth]{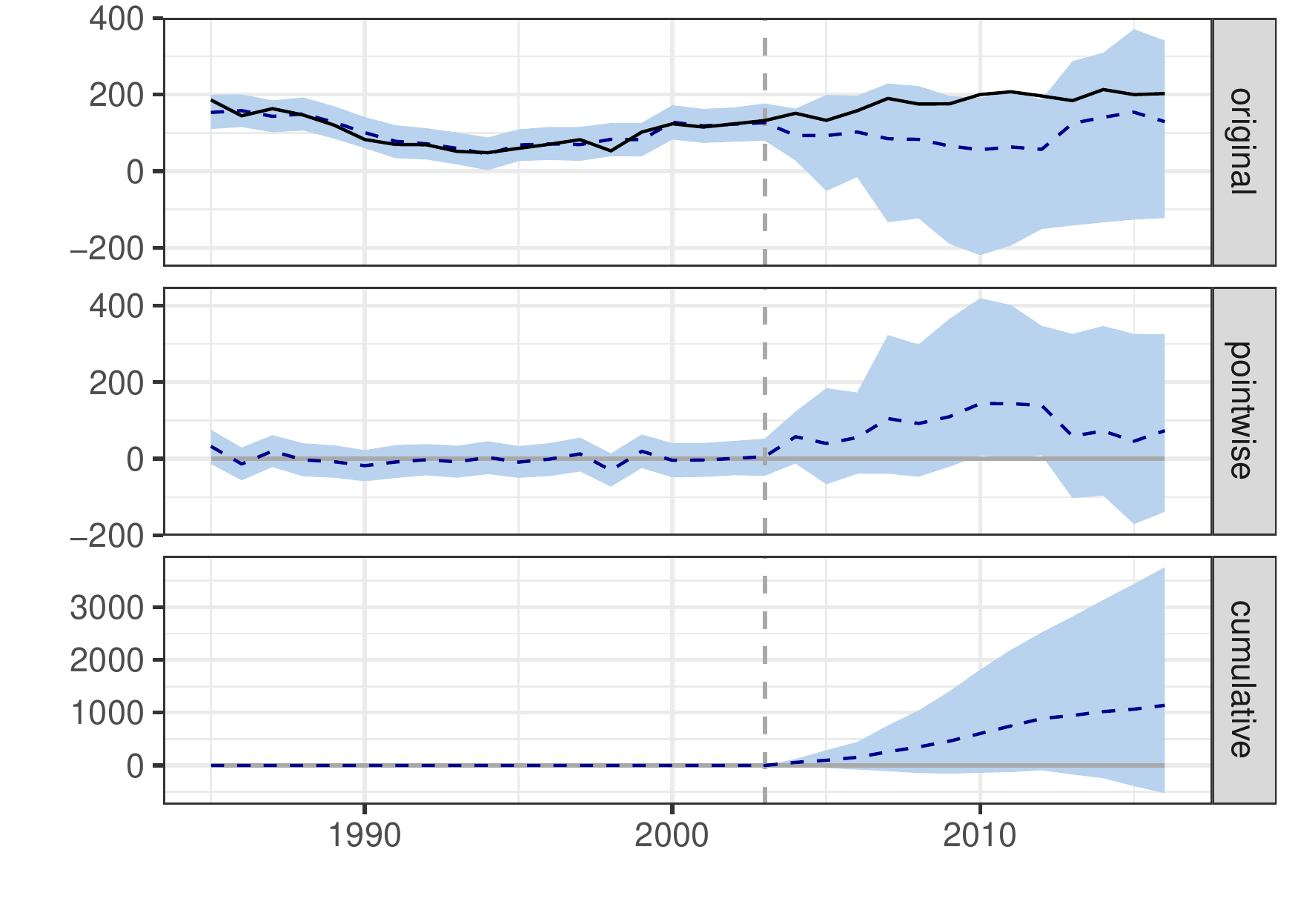}  }
 \hfill
 \subfloat[Latvia]{\includegraphics[width=\mywidth]{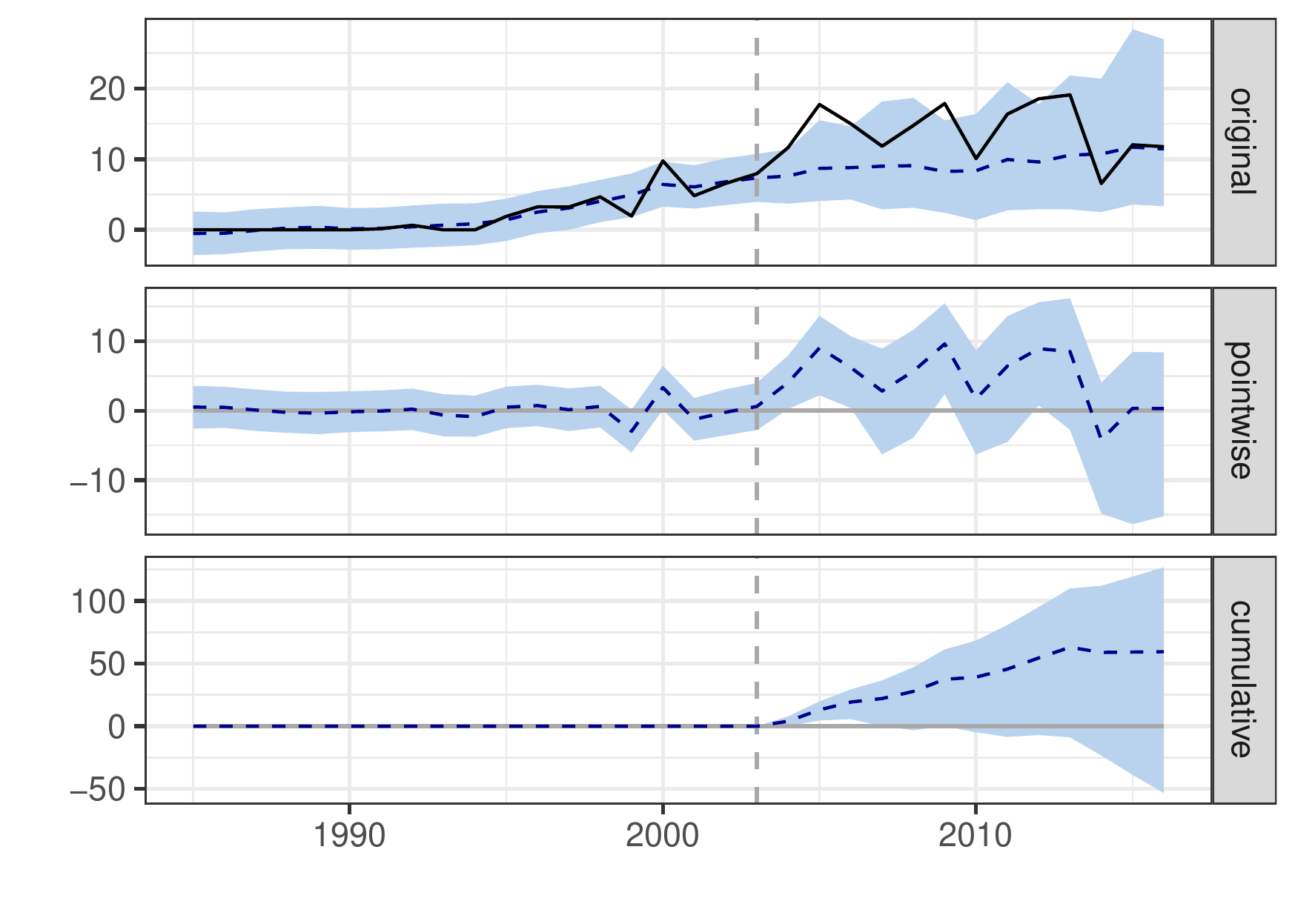}  }\\
 \subfloat[Cyprus]{\includegraphics[width=\mywidth]{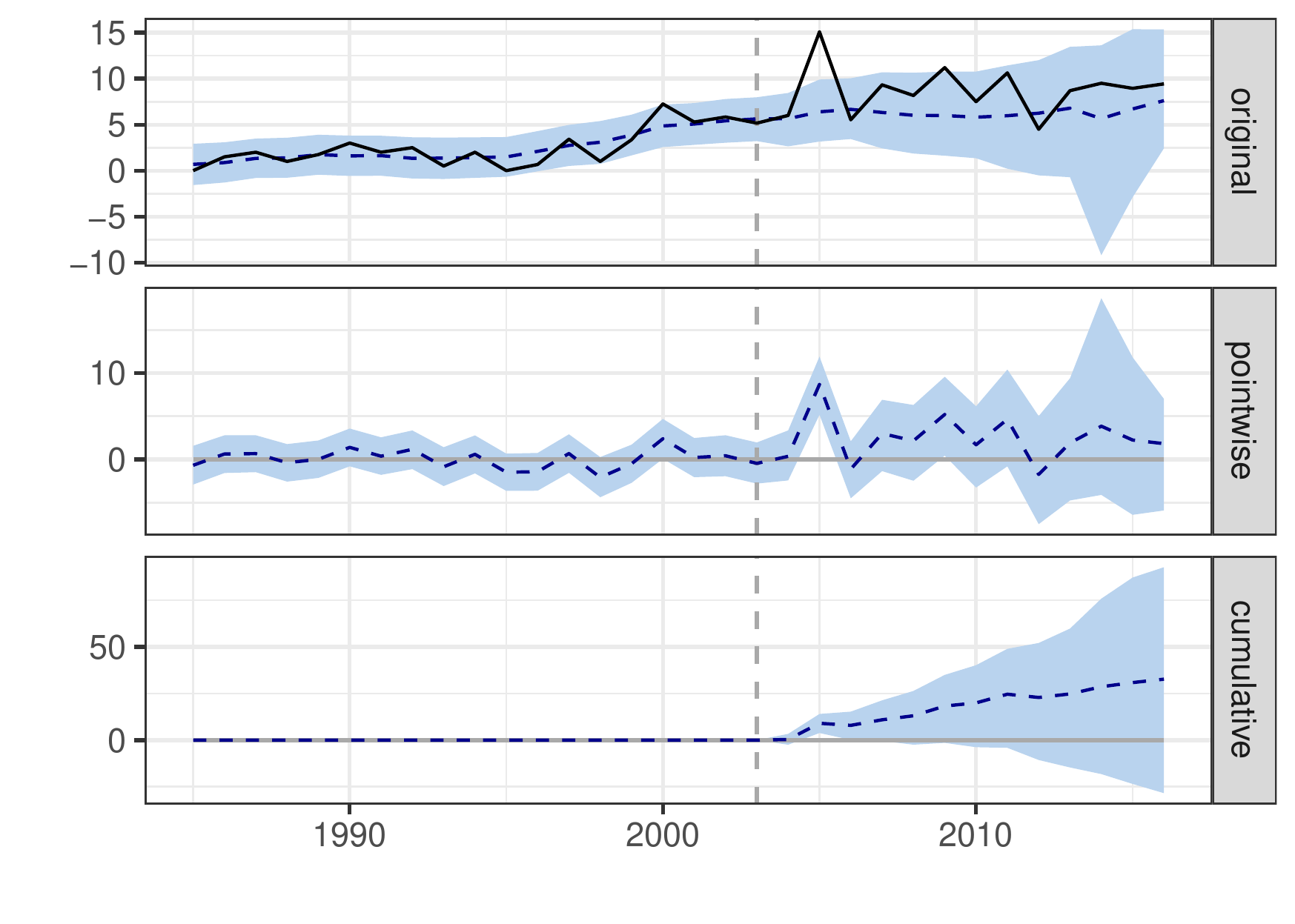}  }
 \hfill
 \subfloat[Slovak Republic]{\includegraphics[width=\mywidth]{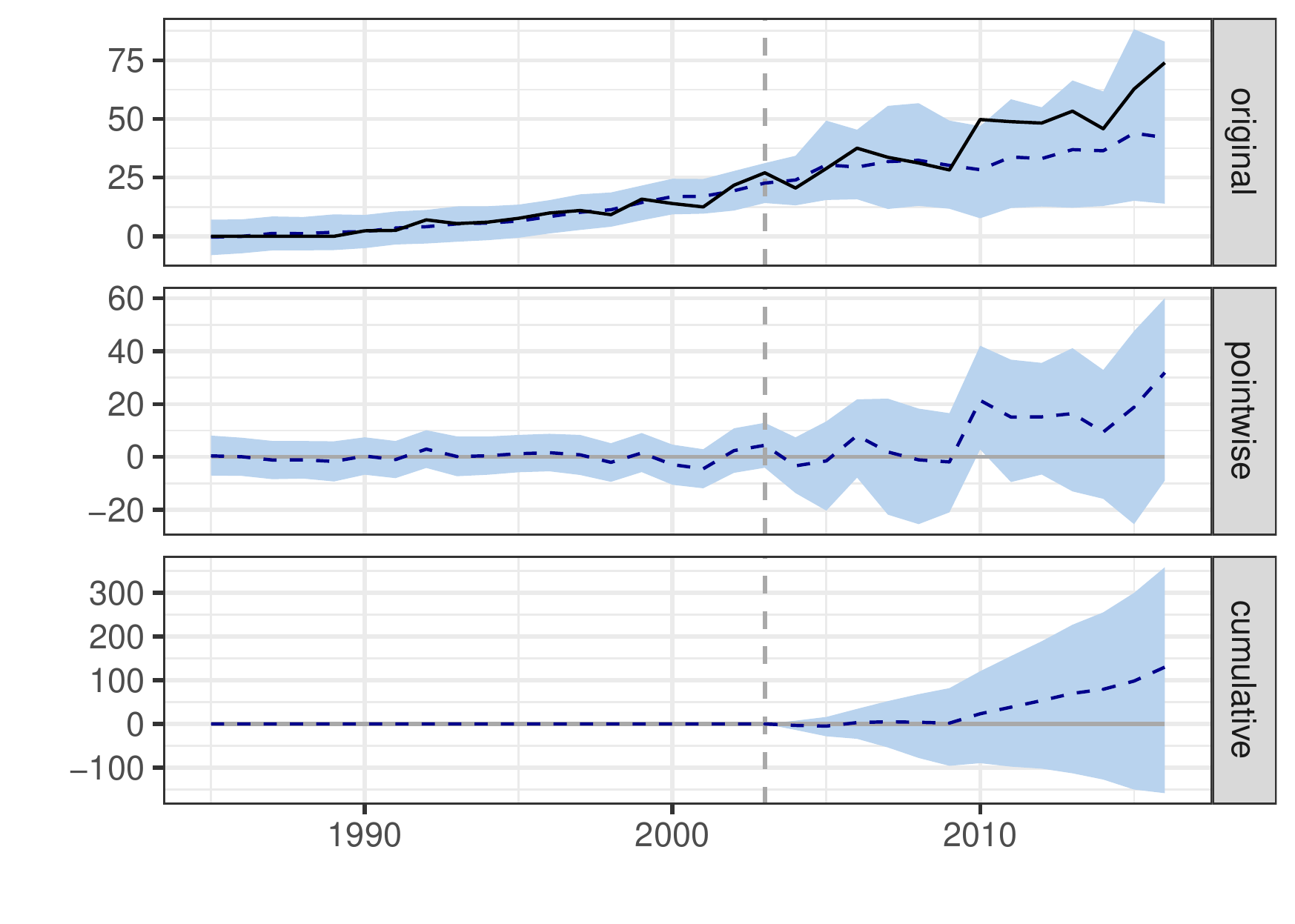}  }\\
 \subfloat[Slovenia]{\includegraphics[width=\mywidth]{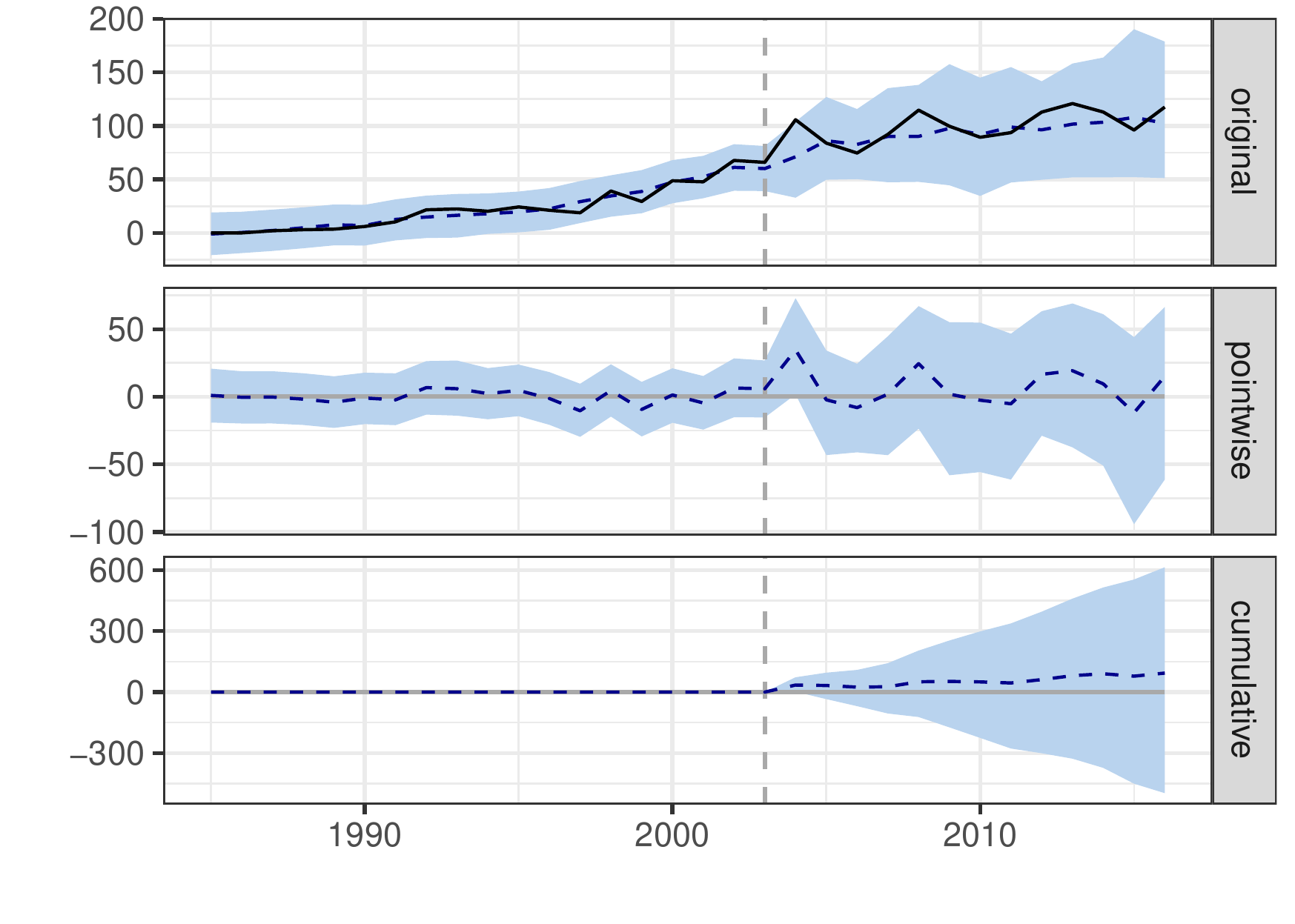}  }
 \hfill
 \subfloat[Bulgaria]{\includegraphics[width=\mywidth]{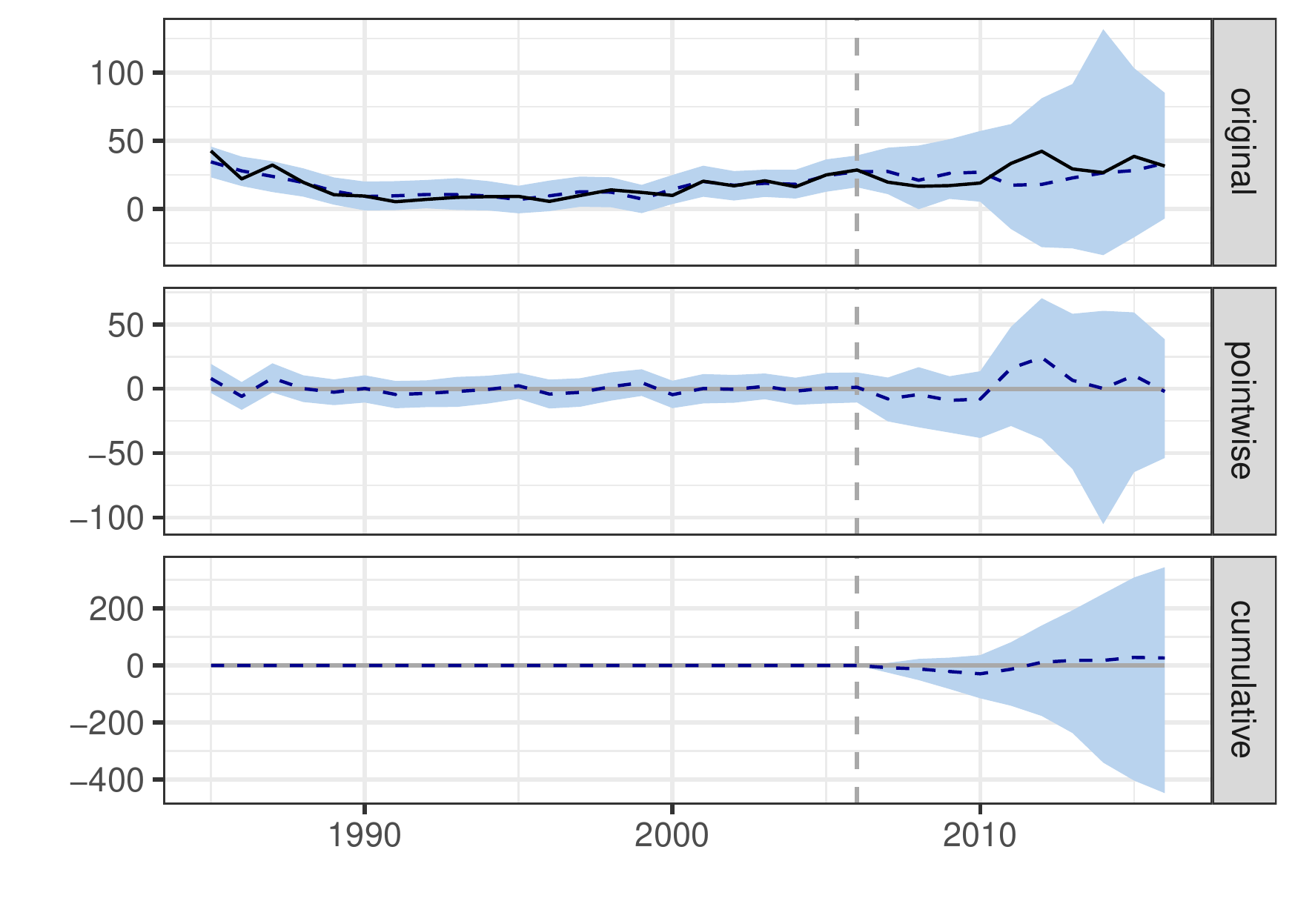}  }\\


    \source{own calculations}
\end{figure}
\begin{figure}\ContinuedFloat
 \caption{Insignificant causal impact (cont)}\label{fig:insigimpactcont}
 \subfloat[Malta]{\includegraphics[width=\mywidth]{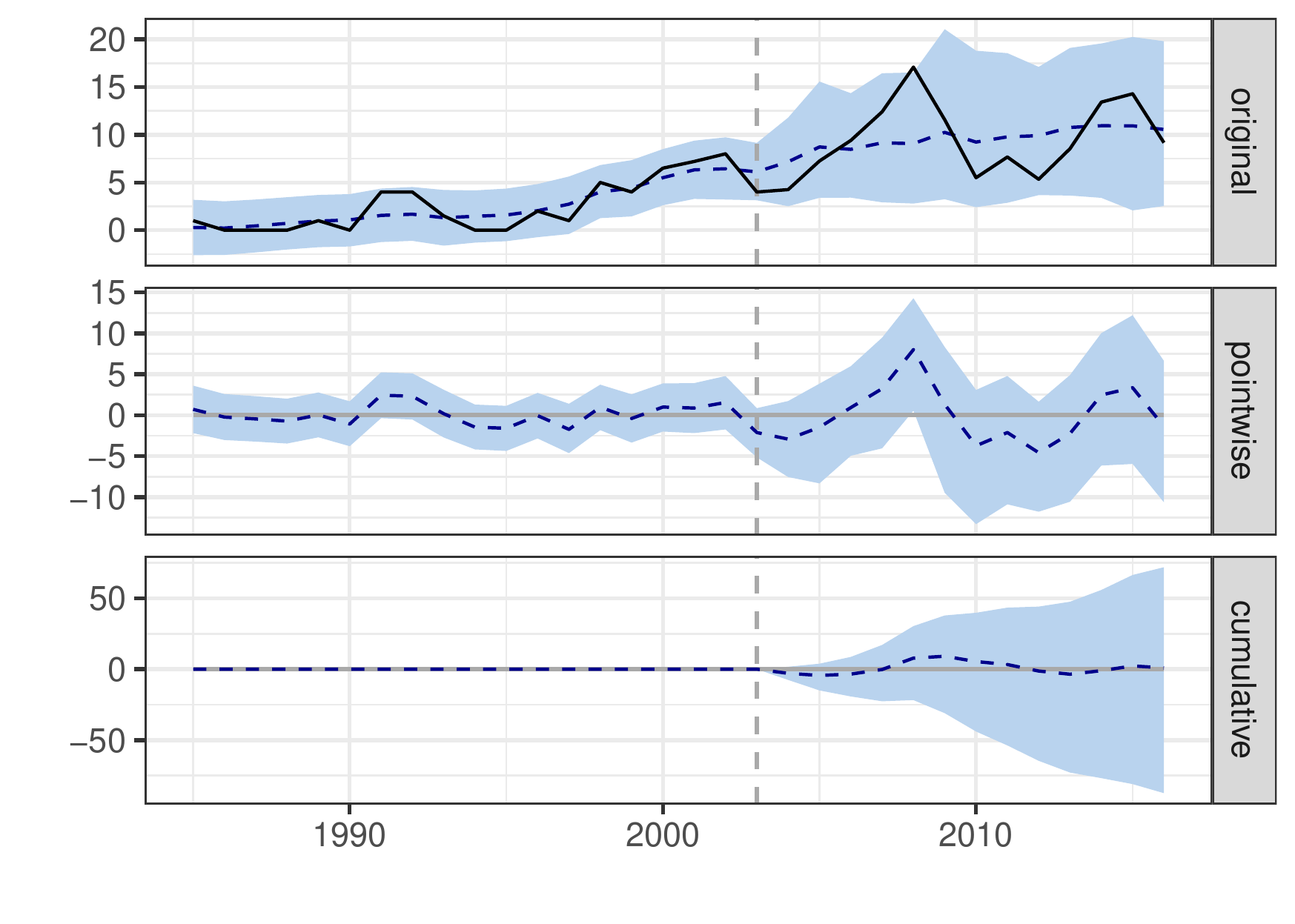}  }


    \source{own calculations}
\end{figure}

\section{Conclusion}
\label{sec:conclusion}

One of the most important initiatives defined by the EU in its Europe 2020 Strategy was to create an innovation-friendly environment that supports the generation, emergence, and diffusion of innovations. Taking it into account, it is worth highlighting that patents play an increasingly important role in innovation and economic performance. In an increasingly knowledge‐driven economy, society invariably needs creative or inventive ideas or concepts to improve existing features, add useful new features to products or develop new ones. Patents are widely used as an indicator of how much innovation is taking place in a given industry. Patents are also one output of technologically successful R\&D activities and a long-term investment. Patenting has experienced a sizeable boom in the last two decades in many fields.
Despite the importance of innovation in the knowledge-driven economy it is also worth highlighting that significant disparities of innovation occur between EU member states, in particular between EU-15 and EU-13. In terms of Summary Innovation Index, the majority of EU-13 are below the EU average. Our analysis also clearly showed that the total number of patents in the years 1985-2017 among the EU member states varies greatly between countries - up to few orders of magnitude. From the analysis a manifestation of the Pareto principle is observed i.e. most patents are issued by only a few countries. Accordingly, the greatest number of patents was assigned in Germany, followed by France, United Kingdom and Italy. Although, EU Membership could bring benefits for member states (e.g. boosted trade in goods or non-tariff trade cost) there are different views on the EU and its influence on member countries. Hence, it is crucial to have a reliable and quantitative assessment of how much the membership influences member states. Influence of EU enlargement with causal  analysis on innovations could be identified in very few scientific publications. In our research we have examined the dynamics of patents performance in the EU-13 over time, in particular the impact of EU-13 accession on their patents performance. The proposed approach, based on causal impact using a Bayesian structural time-series model, indicates a conclusion that joining the EU has brought a statistically significant impact on patents performance in Romania, Estonia, Poland, Czech Republic, Croatia and Lithuania. Having said that, it is worth noting that in Croatia and Lithuania the effects were negative. For the remaining countries: Hungary, Latvia, Cyprus, Slovakia, Slovenia, Bulgaria and Malta we did not find any evidence that joining to the EU has significantly affected their patents performance.

To achieve a high patents performance, countries need an innovation system based on a research system, human resources as well as finance and support. EU member states contribute to the Union’s common budget, but member states also receive money back in various forms. However, EU-13 are less effective in getting research funding (e.g. Horizon 2020 or European Research Council) which could also affect patenting. The presented results could be a valuable input for a policy-oriented discussion on innovation, particularly related to patents performance in each EU country. 
The limitation of our analysis is the dataset because data were available only until 2017. A more enhanced database could enable a causal impact model of the changes of patents cooperation between EU countries after and before accession. This could provide further useful information. Nevertheless, the presented approach which utilised Google’s Causal Impact could be widely used for the assessment of many aspects of EU-13 accession or any other political event. Results from our analysis could be a valuable source of assessment of European enlargement on patents performance and provide further helpful information for policymakers in reforming government subsidies concerning innovations in the New Member States of the EU.


\end{document}